\documentstyle[12pt]{article}
\newcommand{\be}{\begin{equation}}
\newcommand{\ee}{\end{equation}}
\newcommand{\beqn}{\begin{eqnarray}}
\newcommand{\eeqn}{\end{eqnarray}}
\newcommand{\beqnn}{\begin{eqnarray*}}
\newcommand{\eeqnn}{\end{eqnarray*}}
\newcommand{\hr}{\hat\rho}
\def\a{\alpha}

\def\g{\gamma}
\def\l{\lambda}
\def\s{\sigma}
\def\vf{\varphi}
\def\vep{\varepsilon}

\begin{document}

\title{Decoherence and thermalization dynamics of a quantum oscillator}
\author{V V Dodonov\thanks{On leave from Lebedev Physical Institute and
Moscow Institute of Physics and Technology, Russia}
\thanks{e-mail: vdodonov@power.ufscar.br}\ ,\ \
S S Mizrahi\thanks{e-mail: salomon@power.ufscar.br}\ \
and A L de Souza Silva\thanks{On leave from
Universidade Federal de Rond\^onia,
Brazil}
\thanks{e-mail: palus@iris.ufscar.br}
\\
Departamento de F\'{\i}sica, Universidade Federal de
S\~ao Carlos,\\
Via Washington Luiz, km 235, 13565-905  S\~ao Carlos,  SP,  Brasil}
\date{}

\maketitle

\begin{abstract}
We introduce the quantitative measures characterizing the rates of
decoherence and thermalization of quantum systems.
We study the time evolution of these measures in the case
of a quantum harmonic oscillator whose relaxation is described
in the framework of the standard master equation,
for various initial states (coherent, `cat', squeezed and number).
We establish the conditions under which the true decoherence measure
can be approximated by the linear entropy $1-\mbox{Tr}\hat\rho^2$.
We show that at low temperatures and for highly excited initial states
the decoherence process consists of
three distinct stages with quite different
time scales. In particular, the `cat' states preserve $50\%$ of the
initial coherence for a long time interval which increases logarithmically
with increase of the initial energy.

\end{abstract}
%PACS: {03.65.Bz, 42.50.Ar, 42.50.Dv}

\newpage

\section{Introduction}

Recently, a significant interest to the decoherence processes
in quantum mechanics is observed, in particular, due to the
problem of stability of quantum superpositions
(frequently modeled by some kinds of the `Schr\"odinger cats'
%\cite{Scat,DMM-74,YuS,BuK,Brif})
\cite{Scat}-\cite{Brif})
under the influence of the environment
%\cite{Cal,Walls,Sav,Buz,KimBuz,Zur,GGH,Dav,Kim,Miled}.
\cite{Cal}-\cite{Miled}.
However, despite that the qualitative picture of the phenomenon seems
more or less clear,
there are no unique {\em quantitative\/} measures of the rate of
decoherence or the rate of thermalization.
The decoherence implies a degradation of the quantum interference
effects (manifesting themselves in the existence of quantum superpositions)
due to the interactions with the `external world'.
Since these effects are inherent to the pure quantum states,
while they disappear in quantum mixtures,
it seems natural, on the face of it,
to identify the `degree of decoherence'
with the degree of `impurity' of the
quantum state, expressed in terms of
the `canonical entropy' \cite{Dino}
$S=-{\rm Tr}\left(\hat\rho\ln\hat\rho\right)$ or
in terms of the `linear entropy'
\cite{Zur,Kim,Isar-pre}
$s=1-{\rm Tr}\hat\rho^2$,
which is more simple for calculations.
However, a deeper analysis shows that such an identification leads
to certain difficulties, especially in the low temperature case.

Indeed, let us consider the
evolution of an initial {\it pure\/} ($s(0)=S(0)=0$) quantum state
due to a weak interaction with a large reservoir at low temperature.
 For $t>0$, $s(t)$ and $S(t)$ assume positive values,
so the rate of increase of $s(t)$ or $S(t)$ at $t \to 0$
can provide us some hints to the time scale of the initial phase
of the decoherence process \cite{Zur,Kim}.
However, tracing the evolution of the entropies for the long time
interval, we discover that for a small enough temperature of the
environment, the entropies, after reaching some maxima, finally decrease
to very small values which tend to zero when $T\to 0$
(because the stationary {\em thermal mixed state\/} is very close in this
case to the {\em ground pure state\/}).
Then, identifying the measure of quantum impurity
with the measure of decoherence,
one should accept a strange result that the degree of
decoherence of the final {\it equilibrium\/} state is almost the same
(close to zero) as it was initially,
despite that the thermal states are usually believed to be the most
`incoherent'

This example shows that at low temperatures the entropies can serve
as the measures of
decoherence only at the initial stage of the decoherence process.
Thus, several questions arise. The first one: is it possible to
find some other measures which could be used in the whole interval
$0\le t <\infty$ and for any temperature
of the environment? Another question is: under which conditions
(at which time scale) the usage of the entropies as the measures of the
decoherence can be justified?

In the present paper, we answer both questions, introducing
the new parameter ${\cal C}$ whose connection with the degree of coherence
is indubitable (see section \ref{sec-2}).
This parameter equals one for pure quantum states
and $0$ for the thermodynamical equilibrium states,
for any temperature $T>0$.
In section \ref{sec-3}
we introduce another parameter ${\cal D}$, which can be considered as the
`measure of thermalization', since it equals zero for any pure state and
$1$ for the thermodynamically equilibrium state of any quantum system with
an equidistant energy spectrum, for any temperature $T>0$.

Following the time evolution of the parameters ${\cal C}$ and ${\cal D}$
in the process of the thermal relaxation of various initial
states (Fock's, coherent, squeezed, `cat') of the harmonic oscillator,
described in the framework of the `standard master equation'
(sections \ref{sec-4} and \ref{sec-5}),
we find the conditions, under which the linear entropy
can serve as a reasonable `measure of decoherence'.
Moreover, in section \ref{sec-6} we demonstrate that
the decoherence of highly excited initial states at low temperatures
goes through three distinct stages, characterized not by some unique
`decoherence time', but at least by {\em two\/} times with quite
different dependences on the initial energy and the temperature.

The first time $t_1$ (which is usually identified with the time of
decoherence)
is, roughly speaking, inversly proportional to the product of the
energy of quantum fluctuations by the number of photons per mode of
the reservoir. During this short time interval the parameter ${\cal C}$
rapidly decreases from $1$ to some {\em finite\/} value which depends
on the initial state. Then ${\cal C}(t)$ remains at a more or less constant
level (for the `cat' states) or even can increase with time (for the
`squeezed' states). And only after the `ultimate decoherence time' $t_d$,
which
{\em increases\/} (logarithmically) with the increase of the initial energy,
the coherence coefficient goes monotonously to the final zero value.

The evolution of the `thermalization parameter' ${\cal D}(t)$
is similar to certain
extent to the behaviour of $1-{\cal C}(t)$: the initial rapid increase
from zero to some intermediate value, then some stabilization or even
decrease, and the final
rapid transition to the equilibrium unit value after the `thermalization
time' $t_T$,
which also depends logarithmically on the initial energy.
The difference between $t_d$ and $t_T$ consists in their temperature
dependences: $t_d$ has a finite limit value when the temperature $T$ tends
to zero, whereas $t_T$ is inversly proportional
to the absolute temperature at $T\to 0$, in accordance
with the third law of thermodynamics (the inattainability of the absolute
zero of temperature implies that the rate of the relaxation processes
must go to zero as $T\to 0$).

\section{The measure of `coherence' }\label{sec-2}

The controversies of the identification the
decoherence measure with the von Neumann' or linear entropies take their
origin in the invariance of these entropies with respect to the
choice of the basis in the Hilbert space: the entropies do not distinguish
the equilibrium state (or other stationary states in the case of `colored'
or `squeezed' reservoirs) from any other mixed one.
But such a symmetry
with respect to the choice of the basis in the Hilbert space
is obviously broken in the relaxation processses, when all possible
initial states tend to the unique equilibrium state, whose
density matrix is diagonal in the distinct basis formed by the energy
eigenstate of the Hamiltonian (or by some other distinct basis in the
case of more sophisticated artificial reservoirs considered recently).
Thus it seems natural to suppose that
the measure of decoherence must depend explicitly on this distinct basis
$|n\rangle\langle n|$
(the concept of the broken symmetry of the Hilbert space was used
as a basis for introducing the {\em polarized distance\/} between different
quantum states in \cite{dist}).
In some special cases, when it is known exactly to which specific family of
quantum states (e.g., coherent state, even/odd coherent state, etc.)
the initial quantum state belongs, the prefered basis may be different
from the energy one, so that some special approaches can be used, as well.
We shall discuss such a situation later on.
However, in the generic case, when the type of the quantum state is not known
beforehand, the only available information is contained in the set of
matrix elements of the statistical operator with respect to the physically
distinguished energy basis:
$\hat\rho(t)=\sum_{mn} \rho_{mn}|m\rangle\langle n|$. Therefore, it seems
reasonable to define the measure of coherence in terms of the
coefficients  $\rho_{mn}$.
Since the decoherence is usually identified with the disappearence of the
{\em off-diagonal elements\/} of the density matrix $\Vert \rho_{mn}\Vert$,
it is natural to define the normalized {\em measure of coherence\/} as
\be
{\cal C}(t)=\sum_{m\neq n}\left|\rho_{mn}\right|^2(t)/
\sum_{m\neq n}\left|\rho_{mn}\right|^2(0).
\label{def-C1}
\ee
Then ${\cal C}(0)=1$, while ${\cal C}\equiv 0$
for any `completely incoherent' state without off-diagonal matrix
elements in the energy basis (provided initially at least one off-diagonal
element was different from zero).
Introducing the `diagonal part' of the operator $\hr$
\be
\hat\rho_d = \sum_n p_n |n\rangle\langle n|, \quad
p_n \equiv \langle n| \hat\rho |n\rangle ,
\label{def-rhod}
\ee
and taking into account the property
$\mbox{Tr}\left(\hr\hr_d\right)=\mbox{Tr}\left(\hr_d^2\right)$
one can rewrite (\ref{def-C1}) in the form
\be
{\cal C}=\frac{\mbox{Tr}\left[\hat\rho(t) -\hat\rho_d(t)\right]^2}
{\mbox{Tr}\left[\hat\rho(0) -\hat\rho_d(0)\right]^2} \equiv
\frac{\mu(t)-\l(t)}{\mu(0)-\l(0)}
\label{def-C}
\ee
\be
\mu\equiv \mbox{Tr}\hat\rho^2, \quad
\l \equiv \mbox{Tr}\hat\rho_d^2=\sum_n p_n^2
\label{lam-sum}
\ee
We shall call $\mu$ the `total purity' and $\l$ the `diagonal purity'.
In many cases of practical interest
both `purities' can be calculated rather easily.
For example, if one knows the {\em Wigner function\/} \cite{Wig1,Wig2}
(we assume $\hbar \equiv 1$)
\be
    W(q, p)
      =\int d v ~e^{i pv}
      \langle \left. q- \frac{ v}{2}
      \right| \hat{\rho} \left| q
      + \frac{ v}{2} \right. \rangle
\label{defWig}
\ee
\[
{\rm Tr}\hr =\int W(q,p)dqdp/(2\pi)=1
\]
then
\be
\mu =\int W^2(q,p)dqdp/(2\pi).
\label{mu-int}
\ee
As to the `diagonal purity',
it can be calculated either by means of a direct summation of the
series in (\ref{lam-sum}), or, equivalently,
it can be expressed as the integral
\be
\l(t)=\int_0^{2\pi} \frac{d\varphi}{2\pi}\left|G\left(e^{i\varphi};t
\right)\right|^2,
\label{calcmud}
\ee
where
\be
G(z;t)\equiv \sum_{n=0}^{\infty} p_n(t)z^n
\label{def-G}
\ee
is the {\em diagonal generating function\/}.
Below we use both methods.

\section{The measure of thermalization}\label{sec-3}

A qualitative measure of {\em thermalization\/}
can be introduced in the following way.
The analysis of the low temperature behaviour of the entropies shows that
the troubles mentioned in the introduction arise due to
the double nature of the ground state, described by the
density operator $\hat\rho_0\equiv |0\rangle\langle 0|$.
On one hand, this state is {\it pure\/}, with ${\rm Tr}\hat\rho_0^2=1$.
On the other hand, it is the limit of the {\it equilibrium states\/},
which are conceived to be completely decoherent.
Therefore it seems reasonable to exclude
the state $\hat\rho_0$ in some way.
One of the possibilities is to take a simple expression for the
linear entropy and to divide it by a proper time-dependent
factor which would ensure a nonzero limit at $t\to\infty$.
This goal can be achieved, for instance, if one
chooses as the normalizing factor
the Hilbert-Schmidt distance between the states
$\hat\rho(t)$ and $\hat\rho_0$.
If the system under study has a finite number of energy levels (e.g., spin
systems), then there are some grounds to treat the state with the maximal
energy $\hat\rho_f\equiv |E_{max}\rangle\langle E_{max}|$ on the same
footing as $\hat\rho_0$. Thus we arrive at the parameter
(introduced for the first time in \cite{SaoLor}, but identified
erroneously with the measure of decoherence)
\beqn
{\cal D} &=&\frac{1-{\rm Tr}\hat\rho^2}
{\left[
{\rm Tr}\left(\hat\rho-\hat\rho_0\right)^2
{\rm Tr}\left(\hat\rho-\hat\rho_f\right)^2
\right]^{1/2}}
\nonumber\\
&=&\frac{1-\mu }{\left[ \left( 1+\mu -p_{f}\right) \left( 1+\mu
-p_{0}\right) \right] ^{1/2}},
\label{D}
\eeqn
where
 $p_0\equiv {\rm Tr}\left(\hr\hr_0\right)=\langle 0|\hat\rho|0\rangle$
is the ground state occupation probability, while
$p_{f}\equiv {\rm Tr}\left(\hr\hr_f\right)$ is the occupation probability
of the level with the maximal energy
(evidently, for quantum systems with infinite dimensional Hilbert spaces,
such as a harmonic oscillator, $p_f \equiv0$ for any physical
state possessing finite energy).

For systems with {\em equidistant spectra\/}, $E_{n+1}-E_n=\Delta E=const$,
the equilibrium occupation probabilities read
\[
p_{n}=\xi ^{n}(1-\xi )/\left( 1-\xi ^{M}\right)
\]
where $M$ is the total number of levels, $ n=0,1,\ldots ,M-1$
and $\xi=\exp(-\beta\Delta E) <1$ is the Boltzmann factor. Then
\[
\mu _{eq}=(1-\xi )\left( 1+\xi
^{M}\right) /\left[ (1+\xi )\left( 1-\xi ^{M}\right) \right] ,
\]
\[
p_{0}^{(eq)}=(1-\xi )/\left( 1-\xi ^{M}\right)
\]
\[
p_{f}^{(eq)}=\xi
^{M-1}(1-\xi )/\left( 1-\xi ^{M}\right)
\]
and we see that ${\cal D}_{eq}\equiv 1$
{\em for any value\/} $0<\xi <1$, while ${\cal D}\equiv 0$ for all pure
states. For this reason, we may consider ${\cal D}$ as `the measure of
thermalization'.
Note that we have an indeterminacy in (\ref{D}) if $\xi=T= 0$,
thus the case $T=0$ must be excluded.
But as we know the limit of exact zero temperature is an idealization,
thus we do not have to worry on this issue.

For systems with nonequidistant spectra the value of ${\cal D}_{eq}$
depends on temperature, nonetheless the limits at $T\to 0$ and $T\to \infty$
still equal $1$. For $T\to\infty$ we have
$p_0=p_1=\ldots=p_{f}=1/M$, consequently $\mu=M\cdot(1/M)^2=1/M$
and ${\cal D}=1$.
In the low
temperature case $T\to 0$, the equilibrium statistical operator is close to
 $p_0|0\rangle\langle 0|+p_1|1\rangle\langle 1|$ with $p_1\ll 1$
 (where $|1\rangle$ is the first excited state), while the contribution
 of other states can be neglected (we consider the systems with discrete
 energy spectra). Then $p_0+p_1=1$,
 $\mu=p_0^2+p_1^2$, $1+\mu-2p_0=2p_1^2$, $1+\mu-2p_f=2$, and
 $1-\mu= 2p_1$ (up to higher order terms). As a result,
 we have ${\cal D}=1$ again.

\section{Decoherence dynamics}\label{sec-4}
\subsection{Time evolution of the quantum state}

We confine ourselves to the analysis of the time dependence
of the `coherence coefficient' ${\cal C}$ (\ref{def-C})
in the process of
thermal relaxation of the {\em harmonic oscillator\/}
described in the framework of the standard master equation
\cite{Weid,Scul}
(more general models were considered, e.g., in \cite{Isar-pre})
\beqn
&&d\hr/dt =
\g\left(1+\nu\right)\left(2\hat a\hr \hat a^{\dag} -
\hat a^{\dag}\hat a\hr -\hr\hat a^{\dag}\hat a\right)\nonumber\\
&&+\g\nu\left(2\hat a^{\dag}\hr \hat a -
\hat a\hat a^{\dag}\hr -\hr\hat a\hat a^{\dag}\right)
-i\left[\hat a^{\dag}\hat a,\hr\right].
\label{mast}
\eeqn
Here $\hat a$ and $\hat a^{\dag}$ are the usual bosonic
annihilation and creation operators, $\nu$ is the equilibrium
mean number of quanta
in the reservoir corresponding to the given mode,
and $\gamma>0$ is a damping coefficient ($\hbar=\omega=1$).

An immediate consequence of equation (\ref{mast}) is the universal expression
for the purity loss rate in an initial pure
state $\hr^2=\hr=|\psi\rangle\langle \psi|$
(cf. \cite{Zur,Isar-pre,Gal,Isar99})
\be
\left.\dot\mu\right|_{t=0}=
\left.2{\rm Tr}\left(\hr\hat{\dot\rho}\right)\right|_{t=0}=
-4\g\left[\nu+(1+2\nu)\s_a\right],
\label{dotmu}
\ee
where
$
\s_a\equiv \left\langle\psi|\hat{a}^{\dag}\hat{a}|\psi\right\rangle-
\left|\left\langle\psi|\hat{a}|\psi\right\rangle\right|^2$.
The `primary' purity loss rate is minimal for
the coherent states with $\s_a\equiv 0$.
In the generic case it is roughly proportional
to the average number of thermal photons in the reservoir and
to the `reduced' {\it energy of quantum fluctuations\/}
in the initial pure state,
\be
{\cal E}_0=\frac12\left[\s_p^{(0)} +\s_q^{(0)}\right]
\equiv \s_a +\frac12,
\label{defE0}
\ee
where
$\s_{q}^{(0)}$ and $\s_{p}^{(0)}$ are the variances
of the quadrature
components $\hat{q}=\left(\hat{a}+\hat{a}^{\dag}\right)/\sqrt2$ and
$\hat{p}=i\left(\hat{a}^{\dag}-\hat{a}\right)/\sqrt2$ in the initial
pure state $|\psi\rangle$.

Due to equation (\ref{dotmu}) the initial evolution of the `purity'
has the linear dependence on time $\mu(t)=1-t/t_1 +\cdots$, where the
`primary purity loss time' equals
\be
t_1= (4\g)^{-1}\left[\nu+(1+2\nu)\s_a\right]^{-1} .
\label{t1}
\ee
Note that some `microscopic' models, based on an
explicit coupling of the system under study with a large reservoir, result
in a {\it quadratic\/} time dependence
$\mu(t)$ at $t\to 0$ \cite{Kim}.
This apparent contradiction is explained by the fact
that the `microscopic' models and the
{\it phenomenological\/} master equations describe the evolution of the
subsystem in different time scales. Actually, the master equation describes
a `coarse-grained' evolution averaged over many periods of the fast
oscillation, so the physical meaning of the limit $t\to 0$ in the case
of the master equation is quite different from the same (formally) limit
in the `microscopic' models.

To calculate the time dependence of the `purity' $\mu$ in the whole interval
$0<t<\infty$ with the aid of formula (\ref{mu-int})
we need the time dependent Wigner function $W(q,p,t)$.
It obeys the Fokker--Planck equation which follows immediately from
(\ref{mast}):
\beqn
\frac{\partial W}{\partial t} &=&
\frac{\partial }{\partial q}\left([\g q -p]W\right)
+\frac{\partial }{\partial p}\left([\g p +q]W\right)
\nonumber\\&&
+\g\left(\nu +\frac12\right) \left(\frac{\partial^2 W}{\partial q^2}
 +\frac{\partial^2 W}{\partial p^2}\right).
\label{eq-W}
\eeqn
The solution to equation (\ref{eq-W}) can be written as
\be
W(q,p;t)=\int {\cal K}(q,p;t|q',p',0)W(q',p';0)dq'dp'.
\label{def-prop}
\ee
The propagator ${\cal K}(q,p;t|q',p',0)$ was calculated by means of
different methods in \cite{Wang,Ag,Hak,Ris}; the
explicit expressions in the case of most general multidimensional
time-dependent quadratic operator (with respect to $q$, $p$,
$\partial/\partial q$, $\partial/\partial p$) in the right-hand side
of the Fokker--Planck equation were given in \cite{167,183}.
In the case involved the general form found in \cite{167} is reduced to
(see appendix)
\beqn
&&{\cal K}(q,p;t|q',p',0)=\left(2\pi\s_* u\right)^{-1}
\exp\Bigg\{-\left(2\s_* u\right)^{-1} \Big[q_t^2+p_t^2
\nonumber\\&&
%\times
+e^{-2\g t}\left(q'^2+p'^2\right)
-2e^{-\g t}\left(q' q_t +p' p_t \right)
\Big]\Bigg\},
\label{prop}
\eeqn
where $\s_*\equiv \nu+\frac12$,
\[
q_t =q\cos t -p\sin t, \quad
p_t =p\cos t +q\sin t ,
\]
and the `compact time' $u$ is given by
\be
u(t)\equiv 1-e^{-2\g t} .
\label{def-u}
\ee

The consequence of the master equation (\ref{mast}) is the closed set of
equations for the diagonal elements of the density matrix in the Fock
(energy) basis
\beqn
\dot p_n &=& 2\g(1+\nu)\left[(n+1)p_{n+1} -np_n\right]
\nonumber\\&&
+2\g\nu\left[np_{n-1} -(n+1)p_n\right] .
\label{pdot}
\eeqn
These equations, in turn, are equivalent to the simple first order
partial differential equation for the diagonal generating function
(\ref{def-G})
\be
\frac{\partial G}{\partial t} = 2\g (1-z)[1+\nu(1-z)]
\frac{\partial G}{\partial z} -2\g \nu(1-z)G
\label{eqG}
\ee
The solution to (\ref{eqG}) reads
\cite{Zeld,Bar,Ar}
\be
G(z,u)
=\frac{1}{1+\nu u(1-z)}
G_0\left(\frac{z+u(1+\nu)(1-z)}{1+\nu u(1-z)}\right)
\label{F(t)}
\ee
where $ G_0(z)\equiv G(z,0)$.
Putting $z=1$ in (\ref{F(t)}) we verify the normalization condition
$G(1,t)\equiv 1$.

\subsection{Initial coherent states}

As the first example we consider
the evolution of the initial coherent state
$|\a\rangle$, $\a\equiv \sqrt{a}\exp(i\phi)$.
Applying the propagator (\ref{prop}) to the initial Wigner function
\[
W^{(coh)}(0)=2\exp\left[-(q-\sqrt{2a}\cos\phi)^2 -(p-\sqrt{2a}\sin\phi)^2
\right]
\]
we obtain
\beqn
W^{(coh)}(q,p,t)&=&2\xi_{\nu}\exp\Bigg\{-\xi_{\nu}\left(
\left[q-\sqrt{2a}e^{-\g t}\cos(\phi -t)\right]^2
\right.\nonumber\\&&\left.
-\left[p-\sqrt{2a}e^{-\g t}\sin(\phi -t)\right]^2\right)\Bigg\},
\label{Wal}
\eeqn
where
\be
\xi_{\nu}(u)  \equiv (1+2u\nu)^{-1}
\label{cohmu}
\ee
coincides with the `total purity': $\mu^{(coh)}=\xi_{\nu}(u)$.

To calculate the `diagonal purity' we use the explicit expression for
the time-dependent diagonal matrix elements in terms of the
Laguerre polynomials \cite{Zeld,Ar}
\be
p_n^{(coh)}=\frac{(u\nu)^n}{(1+u\nu)^{n+1}}
\exp\left[ \frac{a(u-1)}{1+u\nu}\right] L_n \left[
 \frac{a(u-1)}{u\nu(1+u\nu)}\right]
\label{p-Lag}
\ee
Then the sum in (\ref{lam-sum}) is reduced to the known series \cite{Grad}
\be
\sum_{n=0}^{\infty} L_n(x)L_n(y)z^n= (1-z)^{-1}
\exp\left[z\frac{x+y}{z-1}\right] I_0\left[2\frac{\sqrt{xyz}}{1-z}\right]
\label{Lag-Bes}
\ee
($I_0(z)$ is the modified Bessel function),
so we obtain
\be
\l^{(coh)}= \xi_{\nu}(u)
\exp\left(-\eta\right)
I_0\left(\eta\right),
\label{mudcoh}
\ee
\be
\eta=2a(1-u)\xi_{\nu}(u).
\label{def-eta}
\ee
For $a\ll 1$ it is sufficient to take into account the first terms of the
Taylor expansion of the function $e^{-\eta}I_0(\eta)$ to obtain
\be
{\cal C}^{(coh)}(t)\approx \xi_{\nu}^{2}(u)(1-u)=
\mu^2(t) e^{-2\g t}.
\label{Casmall}
\ee
In this case the time dependence of the `purity' has a little in common
with the time dependence of the `coherence'; the same is true even for
$a\sim 1$: see figure \ref{fig-1}.

For the {\em highly excited\/} initial states with $a\gg 1$
the asymptotics of the modified Bessel function,
$I_0(x)\approx (2\pi x)^{-1/2}e^x$ for $x\gg 1$, yields
\[
{\cal C}^{(coh)}(t)\approx \mu^{(coh)}(t)
\left[1 -\frac1{2\sqrt{a\pi}} \left(\sqrt{1+2u\nu}\,e^{\g t}-1
\right)\right].
\]
In this case, the contribution of the diagonal
elements to the total purity is small compared with the
contribution of the off-diagonal terms, therefore
the sum over $m\neq n$ in (\ref{def-C1}) can
be replaced by the sum over {\em all\/} values of $m$ and $n$, and the
correlation coefficient can be approximated by the purity $\mu$.
This is just the
case considered in most of the papers devoted to
the decoherence of initial `macroscopic' quantum states.
However, ${\cal C}\approx \mu$ only under the condition
$\l\ll \mu$. Consequently, the linear entropy $1-\mu$ can be considered as a
measure of decoherence only for $a\gg 1$, and under the additional
restriction $a\xi_{\nu}(u)\exp(-2\g t)\gg 1$,
i.e.,  at the time scale
\be
t\ll t_*\approx(2\g)^{-1}\ln[(a+2\nu)/(1+2\nu)] .
\label{t*}
\ee
For $t\ge t_*$ the identification of the `purity' with the `coherence'
leads to incorrect results as mentioned in the introduction.

The `final decoherence time' $t_d$ can be defined by means of the equation
\be
{\cal C}(t_d)=\beta \mu_{eq}
\label{eq-td}
\ee
where $\beta <1$ is some small number whose choice is a matter of convention
(say, $\beta=0.1$), and $\mu_{eq}=(1+2\nu)^{-1}$ is the equilibrium value
of the `purity' (we take into account that ${\cal C}$ is proportional to
$\xi_{\nu}=\mu^{(coh)}$, according to equation (\ref{mudcoh})).
To solve equation (\ref{eq-td}) for sufficiently small $\beta$ we can use
the asymptotical form of the coherence function at $\g t \gg 1$
(when $u\approx 1$)
\[
{\cal C}^{(coh)}(t)\approx 2a(1+2\nu)^{-2}e^{-2\g t}, \quad
a \gg 1, \quad \g t \gg 1.
\]
Thus we obtain the estimation
\be
t_d^{(coh)}\approx (2\g)^{-1}\ln[2a\mu_{eq}/\beta]
\label{tdcoh}
\ee
which holds for $ a\mu_{eq}\gg 1$.
The evolution of the `total purity' and the
`coherence coefficient' for highly excited initial coherent states at
different temperatures is shown in figure \ref{fig-2}.

\subsection{Initial `cat' states}

Now let us consider the family of
the initial `Schr\"odinger cat' states \cite{BuK,Brif,Buz}
\begin{equation}
|\alpha;\varphi\rangle= {\cal N}
\left(| \alpha\rangle +e^{i\varphi} | -\alpha\rangle\right),
\label{cat}
 \end{equation}
\be
 {\cal N}=
\left(2\left[1+\cos\varphi \exp(-2a)\right]\right)^{-1/2}, \quad
a\equiv |\alpha|^2.
 \label{normcat}
 \end{equation}
The special cases of this family are even ($\varphi=0$) and
odd ($\varphi=\pi$) coherent states \cite{DMM-74},
and the Yurke-Stoler states ($\varphi=\pi/2$) \cite{YuS}.
The Wigner function of the state (\ref{cat}) reads (hereafter we assume 
$\a=\sqrt{a}$ to be real)
\beqn
&&W^{(cat)}(q,p;0)= 4{\cal N}^2\exp\left(-q^2-p^2\right)
\nonumber\\
&&\times
\left[e^{-2a}\cosh(\sqrt{8a} q) +\cos(\sqrt{8a} p +\varphi)\right].
\label{catwig0}
\eeqn
Applying the propagator (\ref{prop}) to this function we obtain
(see also \cite{KimBuz})
\beqn
&&W^{(cat)}(q,p;t)= 4{\cal N}^2\xi_{\nu}
\exp\left[-\xi_{\nu}\left(q^2+p^2\right)\right]
\nonumber\\
&&\times
\left\{\exp\left[-2a(1-u)\xi_{\nu}\right]
\cosh\left[\xi_{\nu}\sqrt{8a(1-u)}\,q_t
\right]\right.
\nonumber\\
&&\left.
+\exp\left[-2a(1+2\nu)u\xi_{\nu}\right]
\cos\left[\xi_{\nu}\sqrt{8a(1-u)}\, p_t +\varphi\right] \right\}
\label{catwig}
\eeqn
where the function $\xi_{\nu}(u)$ was defined in equation (\ref{cohmu}).
Calculating the integral in (\ref{mu-int}) we find the `total purity'
\beqn
&&\mu=2{\cal N}^4 \xi_{\nu}(u) \Big[1 + 4\cos\varphi e^{-2a}
+\cos(2\varphi) e^{-4a} \nonumber\\
&& +
\exp\left[-4a(1-u)\xi_{\nu}\right]+
\exp\left[-4a u(1+2\nu)\xi_{\nu}\right] \Big]
\label{mucat}
\eeqn
The photon distribution function can be written as
\be
p_n^{(cat)}= 2{\cal N}^2 \left[p_n^{(coh)}(a) +
\cos\varphi e^{-2a} p_n^{(coh)}(-a)\right],
\label{pncat}
\ee
where $p_n^{(coh)}(a)$ is given by (\ref{p-Lag}). Calculating again
the sum $\sum p_n^2$ with the aid of formula (\ref{Lag-Bes})
we find
\beqn
&&\l^{(cat)}=4{\cal N}^4\xi_{\nu}(u)\Bigg\{I_0(\eta)
\left[e^{-\eta} + \cos^2\vf e^{\eta-4a}\right]
\nonumber\\&&
+2\cos\vf e^{-2a} J_0(\eta)\Bigg\}
\label{lam-cat}
\eeqn
where $J_0(z)$ means the usual Bessel function, and
$\eta$ was defined in equation (\ref{def-eta}).
If $a\ll 1$, then we have $\mu\approx \xi_{\nu}(u)$ and the same
formula (\ref{Casmall}) for $\lambda$.

For highly excited (`macroscopic': $a\gg 1$) initial cat states the
phase $\vf$ becomes unimportant, and ${\cal N}^2 \approx 1/2$.
Until $a(1-u)\equiv a\exp(-2\g t)\gg 1$,
the `diagonal purity' is small, similarly to the case of coherent
states: $\l^{(cat)}\sim \exp(\g t)/\sqrt{a}$. Then
the coherence coefficient
can be replaced by the `total purity' $\mu$, which rapidly decreases from
 $1$ to the value $\frac12\xi_{\nu}(u)$:
\[
\mu \approx \frac12\xi_{\nu}(u)\Big(1+ \exp
\left[-4a u(1+2\nu)\xi_{\nu}(u)\right]\Big),
\]
staying at this level until $a(1-u)$ becomes smaller than $1$.
In particular, in the low temperature case $\nu\ll 1$ we observe the
`plateau' $\mu\approx{\cal C}\approx \frac12$: see figure (\ref{fig-3}).

At the final stage of the evolution $\mu$ goes to the equilibrium value
$\mu_{eq}=(1+2\nu)^{-1}$ as
\[
 \mu \approx \frac12\xi_{\nu}(u)\Big(1+ \exp[-4a(1-u)\xi_{\nu}(u)]\Big),
\]
but now it becomes compatible with the `diagonal purity'
$\l\approx \xi_{\nu}(u)\exp[-2a(1-u)\xi_{\nu}(u)]$,
so
\[
{\cal C}^{(cat)} \approx \frac12\xi_{\nu}(u)\Big(1-
\exp[-2a(1-u)\xi_{\nu}(u)]\Big)^2.
\]
For $a(1-u)\ll 1$ we have
$ {\cal C}^{(cat)} \approx 2a^2\mu_{eq}^3\exp(-4\g t)$.
Then equation (\ref{eq-td}) yields
the `ultimate decoherence time'
$t_d^{(cat)} \approx (2\gamma)^{-1}\ln\left[a\mu_{eq}\sqrt{2/\beta}\right]$
(if $a\mu_{eq}\gg 1$),
which is only slightly less than the similar time $t_d^{(coh)}$
(\ref{tdcoh}).

\subsection{Decoherence in the accompanying basis}

In the special case of the `cat' states there exists another natural choice
of the `diagonal' part of the statistical operator,
different from (\ref{def-rhod}).
Namely, one can define the `accompanying' diagonal operator as
$\hr_{mix}=\frac12\left(|\a\rangle\langle\a|
+|-\a\rangle\langle-\a|\right)$.
The Wigner function of this quantum mixture is the sum of two coherent
Wigner functions (\ref{Wal}) with opposite values of parameter $\a$:
\beqn
&&W_{mix}(q,p;t)=\frac12\left[W_{\a} +W_{-\a}\right]
\nonumber\\ &&=
2\xi_{\nu}(u)
\exp\left\{-\xi_{\nu}(u)\left[q^2+p^2 +2a(1-u)\right]\right\}
\nonumber\\ && \times
\cosh\left[\xi_{\nu}(u)\sqrt{8a(1-u)}\,q_t \right].
\label{catmix}
\eeqn
Then the `accompanying' (normalized) coherence coefficient can be
defined as
\beqn
&&{\cal F}=\int\left[W^{(cat)}(t)-W_{mix}(t)\right]^2dqdp
\nonumber\\ && \times
\left\{\int\left[W^{(cat)}(0)-W_{mix}(0)\right]^2dqdp\right\}^{-1}
\label{defcalF}
\eeqn
Performing the calculations we obtain
\beqn
&&{\cal F}= \frac{ \xi_{\nu}(u)
\left( 1- \exp\left[-4a(1-u)\xi_{\nu}\right]\right)}
{\left(1 -\cos^2\vf e^{-4a}\right)\left(1 - e^{-4a}\right)}
\nonumber\\ && \times
\left( \exp\left[-4a u(1+2\nu)\xi_{\nu}\right] -\cos^2\varphi e^{-4a}
\right)
\label{Fcat}
\eeqn
 The time evolution of this parameter is essentially different from
 the evolution of the coefficient ${\cal C}$: see figure (\ref{fig-4}).
 This is not surprising,
since the phenomena observed from the moving (`accompanying')
frame of reference in many cases look very different, compared with their
appearance in the fixed frame.

\subsection{Initial squeezed states}

Now let us consider the states
possessing the {\it Gaussian Wigner functions\/}
\cite{Isar99,167,Grab,1mod}
\beqn
&&W(q,p)=d^{-1/2}\exp\Bigg\{-\frac1{2 d}\Big[
\s_{p}(q-\bar{q})^2
\nonumber\\&&
-2\s_{qp}(q-\bar{q})(p-\bar{p})
+\s_{q}(p-\bar{p})^2\Big]\Bigg\},
\label{Wgaus}
\eeqn
where $\s_{q}$, $\s_{p}$ and $\s_{qp}=\s_{pq}$ are the (co)variances
of the quadrature components,
whereas $\bar{q}$ and $\bar{p}$ stand for the average
values of these quadratures.
The parameter
$
 d\equiv \s_{p} \s_{q}- \s_{pq}^2
$
must satisfy the Schr\"odinger--Robertson uncertainty relation
\cite{183,SR,Kurm} $d\ge 1/4$. It is related to the `purity' of the
state as
$ \mu=(4d)^{-1/2}$.
In the thermal state,
$\s_{q}^{(eq)}=\s_{p}^{(eq)}=\s_* \equiv \frac12 +\nu$,
$\s_{pq}=\bar{q}=\bar{p}=0$.

The evolution of the five parameters determining the Gaussian state is
governed by the set of equations following from the master equation
(\ref{mast})
\be
d\bar{q}/dt=\bar{p}-\g\bar{q}, \quad
d\bar{p}/dt=-\bar{q}-\g\bar{p}
\label{dqp}
\ee
\beqn
\dot\s_{q}&=& 2\s_{qp} -2\g\s_{q} +\g (1+2\nu)
\label{equationsqq}\\
\dot\s_{p}&=& -2\s_{qp} -2\g\s_{p} +\g (1+2\nu)
\label{equationspp}\\
\dot\s_{qp}&=& \s_{p} -\s_{q} -2\g\s_{qp}
\label{equationsqp}
\eeqn
The solutions read \cite{167}
\beqn
\bar{q}(t)&=& e^{-\g t}\left[\bar{q}_{0}\cos t +\bar{p}_{0}\sin t\right]
\label{solq}\\
\bar{p}(t)&=&e^{-\g t}\left[\bar{p}_{0}\cos t -\bar{q}_{0}\sin t\right]
\label{solp}
\eeqn
\beqn
&&\s_{q}(t)=\s_* + e^{-2\g t}\left[
\left(\s_{q}^{(0)}-\s_*\right)\cos^2 t
\right.\nonumber\\&& \left.
+ \left(\s_{p}^{(0)}-\s_*\right)\sin^2 t +
\s_{pq}^{(0)}\sin(2t)\right]
\label{solsqq}
\eeqn
\beqn
&&\s_{p}(t)=\s_* + e^{-2\g t}\left[
\left(\s_{p}^{(0)}-\s_*\right)\cos^2 t 
\right.\nonumber\\&& \left.
+ \left(\s_{q}^{(0)}-\s_*\right)\sin^2 t -
\s_{pq}^{(0)}\sin(2t)\right]
\label{solspp}
\eeqn
\be
\s_{qp}(t)= e^{-2\g t}\left[
\s_{qp}^{(0)}\cos(2 t)
%\right.\nonumber\\&&
+ \frac12
\left(\s_{p}^{(0)}-\s_{q}^{(0)}\right)\sin(2 t)\right]
\label{solsqp}
\ee
Any initial {\em pure\/} Gaussian state is unitarily equivalent to the
{\it squeezed\/} state,
defined as an eigenstate of the canonically transformed operator
$\hat b=\cosh\rho\, \hat a +\sinh\rho\, \hat a^{\dagger}$
 with a complex eigenvalue $\a\equiv \sqrt{a}\exp(i\phi)$
and a real `squeezing parameter' $\rho$
(sometimes it is called also a `two-photon state' \cite{Yuen}).
Therefore we parametrize the initial variances and average values as
\[
\s_{q}^{(0)}=\frac12 e^{-2\rho}, \quad
\s_{p}^{(0)}=\frac12 e^{2\rho},  \quad \s_{pq}^{(0)}=0,
\]
\[
\bar{q}_{0}=\sqrt{2 a} e^{-\rho}\cos\phi, \quad
\bar{p}_{0}=\sqrt{2a} e^{\rho}\sin\phi.
\]

The `total purity' $\mu$ does not depend on the first order
moments of the coordinates:
\be
\mu=\left[(1+2u\nu)^2 +4u(1-u)(1+2\nu)\sinh^2\rho\right]^{-1/2}.
\label{musq}
\ee
On the contrary,
the `diagonal purity' $\l$ depends on all 5 parameters of the
one-dimensional Gaussian Wigner function.

The generic diagonal generating function found in \cite{1mod,DM94}
can be expressed as
\be
G(z)=[{\cal G}(z)]^{-1/2}\exp\left(\frac1{D}\left[
\frac{zg_1 -z^2 g_2}{{\cal G}(z)} - g_0\right] \right)
\label{GF}
\ee
where
\[
{\cal G}(z)=\frac14\left[(1+z)^2 +4d (1-z)^2 +
2(\s_q+\s_p)\left(1-z^2\right)\right]
%\label{calG}
\]
\[
D=1+2(\s_p+\s_q) +4 d
\]
\[
g_0= \bar{p}^2(2\s_q+1) +\bar{q}^2(2\s_p+1)
-4\bar{p}\bar{q} \s_{pq}
\]
\begin{eqnarray*}
g_1&=& 2\bar{p}^2\left[\s_q^2 +\s_{pq}^2 +\s_q +\frac14\right]
+2\bar{q}^2\left[\s_p^2 +\s_{pq}^2 +\s_p +\frac14\right]\\
&&-4\bar{p}\bar{q} \s_{pq}(\s_q+\s_p+1)
\end{eqnarray*}
\beqnn
g_2&=& 2\bar{p}^2\left(\s_q^2 +\s_{pq}^2 -\frac14\right)
+2\bar{q}^2\left(\s_p^2 +\s_{pq}^2 -\frac14\right) \\ &&
-4\bar{p}\bar{q} \s_{pq}(\s_q+\s_p)
\eeqnn
For the initial pure squeezed states $G(z,t)$ can be written as
\beqn
&&G(z;t)=\left(f-bz+cz^2\right)^{-1/2}
\nonumber\\&&\times
\exp\left[-a (1-u)\,
\frac{F-Bz+Cz^2}{f-bz+cz^2}\right]
\label{Gt}
\eeqn
where
\[
f=(1+u\nu)^2 +(1 -u)(1+u+2u\nu)\sinh^2\rho,
\]
\[
b=2u\nu(1+u\nu) +2u(1+2\nu)(1-u)\sinh^2\rho,
\]
\[
c=(u\nu)^2 -(1-u)(1-u-2u\nu)\sinh^2\rho,
\]
\[
F=\frac12\left[1-u +R(1+u+2u\nu)\right]
\]
\[
B=1-u +Ru(1+2\nu)
\]
\[
C=\frac12\left[1-u -R(1-u-2u\nu)\right]
\]
\[
R=\cosh(2\rho) -\sinh(2\rho)\cos(2\phi)
\]
Using formula (\ref{calcmud}) and the relations
$$f-b+c=1, \quad F-B+C=0$$
we obtain after some algebra the following integral representation
for the time dependent `diagonal purity' of a generic initially
squeezed state:
\be
\l=\int_0^{\frac{\pi}{2}}\frac{2d\g}{\pi\sqrt{\Phi(\g)}}
\exp\left[-4a(1-u)\left(V+Y\sin^2\g\right)\frac{\sin^2\g}{\Phi(\g)}
\right]
\label{intmud}
\ee
where
\[
\Phi(\g)=\left(1+2b\sin^2\g\right)^2 +(1-u)^2\sinh^2(2\rho)\sin^2(2\g)
\]
\[
V= R(1+2u\nu) +(1-u)\left(R-1 +4R\sinh^2\rho \right)
\]
\beqnn
Y&=&4uB\left[\nu(1+u\nu) + (1-u)(1+2\nu)\sinh^2\rho\right]
\\&&
+2(1-u)\left(1-R -2R\sinh^2\rho \right)
\eeqnn
The evolution of the `purity' and the `coherence coefficient' for the
highly squeezed ($\rho>1$) initial state is illustrated in
figure (\ref{fig-5}). In this case $\mu\approx {\cal C}$ up to the values
of the dimensionless time $\tau=2\g t\sim 1$.
In contradistinction
to the cases of coherent or `cat' states, the coherence coefficient is not
monotonous function of time, but it tries to follow the increase of the
`purity' at $\tau>1$, before going finally to zero.

The integral (\ref{intmud}) can be easily calculated in the long-time
limit $1-u =\exp(-2\g t) \equiv \vep \ll 1$ at zero temperature ($\nu=0$):
$\l= 1-2\vep\left(\sinh^2\rho +aR\right) +{\cal O}(\vep^2)$.
Comparing this expression with the similar expansion of the `total purity'
$\mu$ (\ref{musq}) we obtain
${\cal C}\approx \mu-\l \approx 2\vep aR=2\vep {\cal E}_{cl}$,
where
\be
{\cal E}_{cl}\equiv \frac12\left(\bar{q}_{0}^2 + \bar{p}_{0}^2\right)
=aR
\label{Ecl}
\ee
is the initial `classical' energy (the total energy without the
contribution of the vacuum fluctuations).
Then equation (\ref{eq-td}) yields the `ultimate decoherence time'
$t_d^{(sq)}\sim(2\g)^{-1}\ln\left(2{\cal E}_{cl}/\beta\right)$
which has the same order of magnitude as the time $t_d^{(coh)}$
(\ref{tdcoh}) for the coherent state with the same `classical energy'
$|\a|^2$ (at zero temperature $\mu_{eq}=1$).
If $a=0$ (the initial squeezed
vacuum state), then one should calculate $\mu$ and $\l$ up to the second
order terms with respect to $\vep$. In this case we obtain
${\cal C}\approx \mu-\l \approx \frac14\vep^2\sinh^2(2\rho)$, and
$t_d^{(sq)}\sim(2\g)^{-1}\ln\left[\sinh(2\rho)/2\sqrt{\beta}\right]$.
Since we consider the case $\rho\gg 1$, we can replace $\sinh(2\rho)/2$
by $\sinh^2(\rho)=E-\frac12\approx E$, where $E$ is the {\em total\/}
energy in the case discussed.

\section{Thermalization dynamics}\label{sec-5}

The concrete evolution of the thermalization coefficient depends on
a peculiar
`competition' between the `total purity' $\mu(u)$ (which was calculated
in the preceding sections) and the ground state probability $p_0(u)$,
which can can be easily found from the `diagonal generating function'
\be
p_0(t)=\frac{1}{1+\nu u(t)}
G_0\left(\frac{u(t)(1+\nu)}{1+\nu u(t)}\right).
\label{p0(t)}
\ee
In the case of initial coherent state we have ($a\equiv |\a|^2$)
$G_0^{(coh)}(z)=\exp\left[a(z-1)\right]$. Consequently,
\be
p_0^{(coh)}= \frac1{1+u\nu}\exp\left[-\frac{
a(1-u)}{1+u\nu} \right],
\label{cohp0}
\ee
so the `thermalization coefficient' reads
\[
{\cal D}_a^{(coh)}(u)=
\left\{1 +\frac{1+2u\nu}{(u\nu)^2}\left(1-\exp\left[-
\frac{a(1-u)}{1+u\nu}\right]\right)\right\}^{-1/2}
%\label{D-coh}
\]
Evidently, the case $\nu=0$ should be excluded in this expression.
The dependence on the displacement parameter $a$ disappears
for $a(1-u)\gg 1$, when all the functions ${\cal D}_a^{(coh)}(u)$ merge to
${\cal D}_{\infty}^{(coh)}(u)\approx u\nu/(1+u\nu)$.
This slow evolution is transformed into a fast transition to the equilibrium
value if $a(1-u)\ll 1$:
\be
{\cal D}_a^{(coh)}(1-\varepsilon)\approx 1-
\frac{1+2\nu}{2(1+\nu)}\frac{a\varepsilon }{\nu^2}
+{\cal O}\left(\varepsilon^2\right).
\label{coh-fast}
\ee

For the `cat' states equations (\ref{p-Lag}) and (\ref{pncat}) yield
\beqn
&&p_0^{(cat)} = \frac{2{\cal N}^2}{1+u\nu}
\exp\left[-\frac{a(1-u)}{1+u\nu}\right]
\nonumber\\&& \times
\left\{ 1+ \cos\varphi \exp\left[-\frac{2a u(1+\nu)}
{1+u\nu}\right]
\right\}.
\label{catp0}
\eeqn
At low temperatures ($\nu\ll 1$) all the exponential functions `die out'
if $a(1-u)/(1+\nu)\gg 1$ and $a u/(1+\nu)\gg 1$,
and we observe the `plateau'
${\cal D}^{(cat)}(u)\approx \frac13$ (see figure \ref{fig-6}).

For the squeezed state, formula (\ref{Gt}) yields
\be
p_0^{(sqz)}=G(0,t)=f^{-1/2}\exp\left[-a (1-u)F/f\right].
\label{sqzp0}
\ee
For large values of the squeezing parameter $\rho$
and $\nu\ll 1$, the `purity' (which does not depend on $a$)
is given by
$\mu \approx \left(2\sinh\rho\sqrt{u(1-u)}\right)^{-1}\ll 1$, unless
$u$ is close enough to $0$ or $1$.
If $a(1-u)\gg 1$, then $p_0\ll 1$, and
we observe a universal
(independent of $\a$) behavior of the thermalization coefficient
\[
{\cal D}_{\infty}^{(sq)}(u)\approx \frac{1-\mu}{1+\mu}\approx
\frac{4u(1-u)\sinh^2 \rho}
{\left(\sqrt{1+ 4u(1-u)\sinh^2 \rho} +1\right)^2}
\]
The ${\cal D}$-factor rapidly increases for a small time interval
$t<t_1\sim (\g\sinh^2\rho)^{-1}$ (note that $\sinh^2\rho$ is just
the `reduced' energy of {\it fluctuations\/} in the initial
squeezed state, ${\cal E}_0-\frac12$). For $t>t_1$ we observe some
`plateau', whose extension corresponds approximately to
the interval $0.07<u<0.93$ (inside this interval, the values of
the function $u(1-u)$ are not less than a half of the maximum value
at $u=0.5$).
For the values of $u$ close to $1$, the
${\cal D}$-factor may decrease, following the decreasing
 linear entropy, but finally the term $p_0$ enters the game and
 prevents the thermalization coefficient from falling down to zero.
This final stage of evolution seems very fast in terms of the
`compact time' $u$, but it is not so dramatic with respect to the
scaled time $\tau=2\g t$: see figure \ref{fig-7}.

\begin{figure} \caption
{The `thermalization coefficient' ${\cal D}$
versus the `compact time' $u\equiv 1-e^{-2\gamma t}$ for the initial
{\em coherent\/} and {\em Fock\/} states with equal mean numbers of photons:
$M=|\alpha|^2=1$ and $M=|\alpha|^2=20$, in the low temperature ($\nu=0.01$)
and high temperature ($\nu=10$) cases.
I -- the Fock state with $M=1$ and $\nu=10$,
II -- the Fock state with $M=20$ and $\nu=10$,
III -- the Fock state with $M=20$ and $\nu=0.01$,
IV -- the coherent state with $|\alpha|^2=20$ and $\nu=10$,
V -- the Fock state with $M=1$ and $\nu=0.01$,
VI -- the coherent state with $|\alpha|^2=1$ and $\nu=0.01$,
VII -- the coherent state with $|\alpha|^2=20$ and $\nu=0.01$.
The internal box shows the behaviour of the last two curves in another
scale.
\label{fig-8}
}
\end{figure}
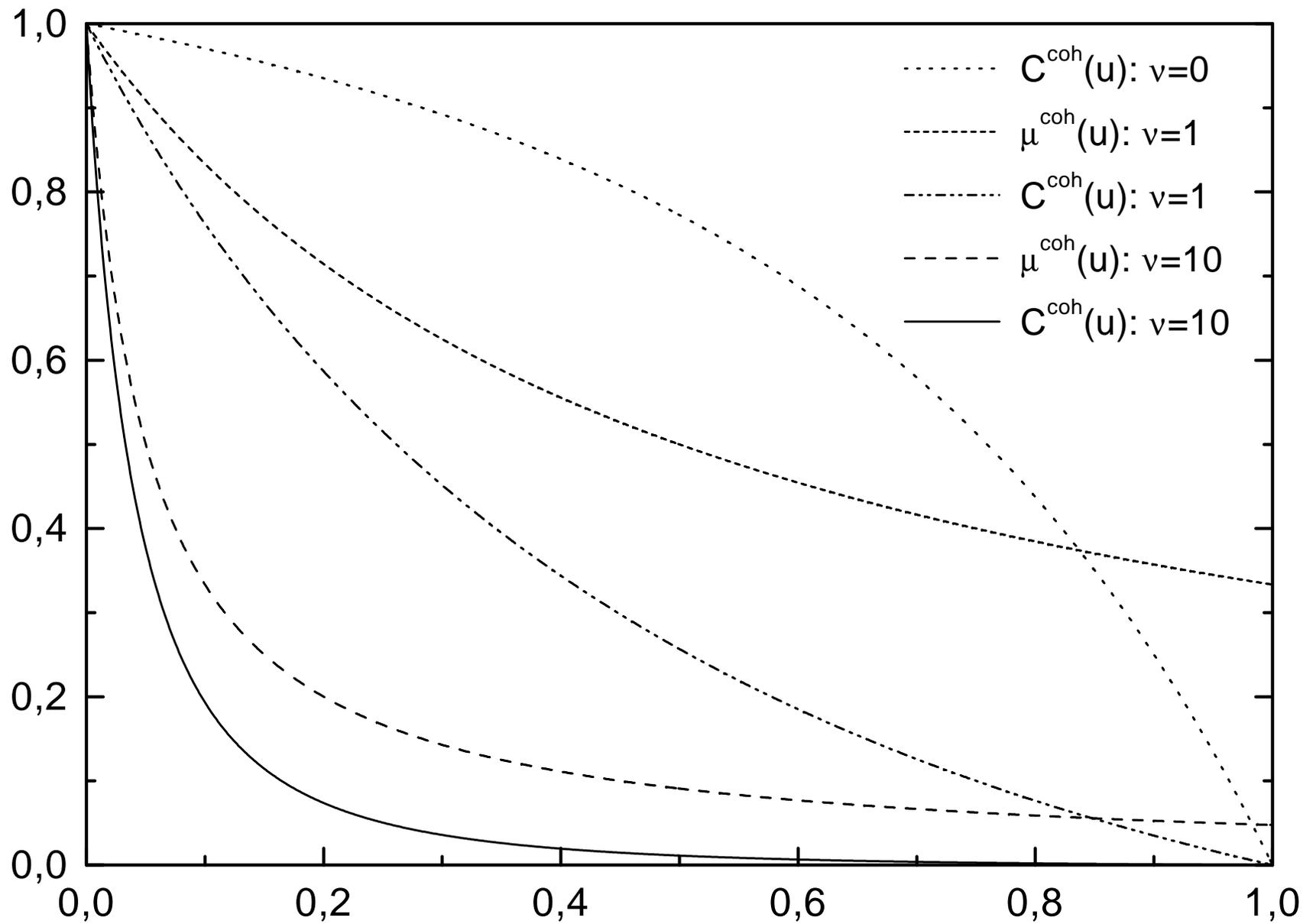
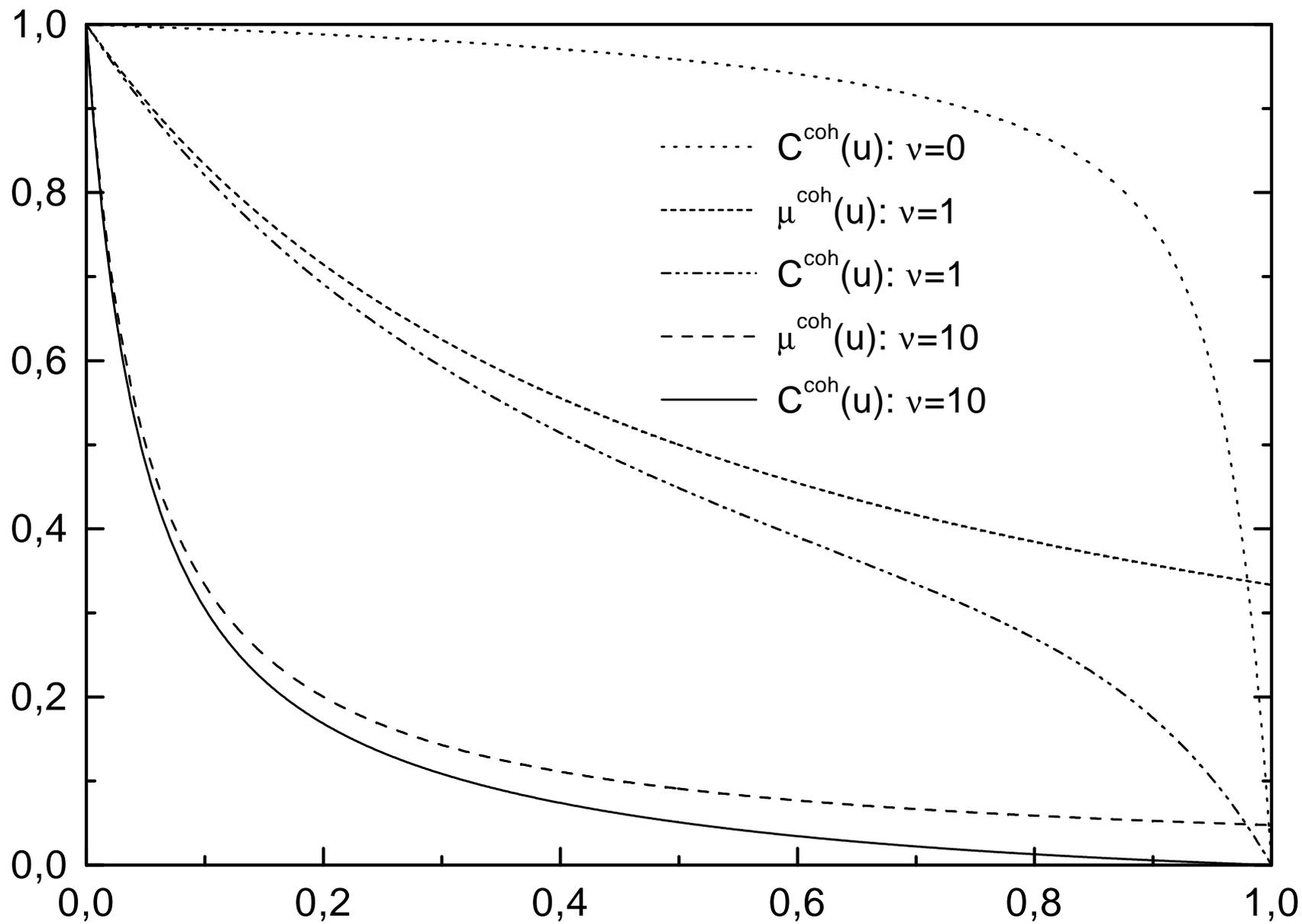
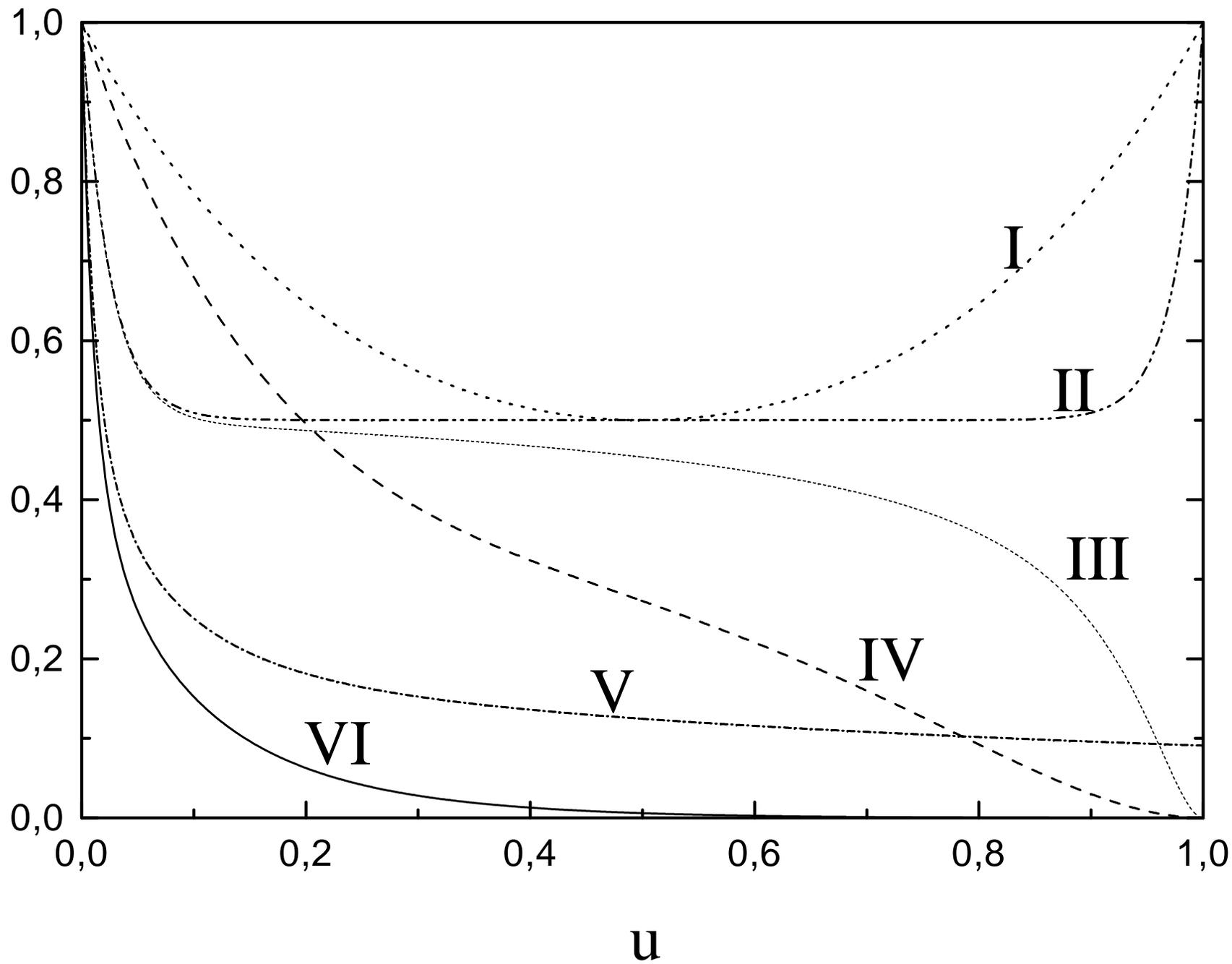
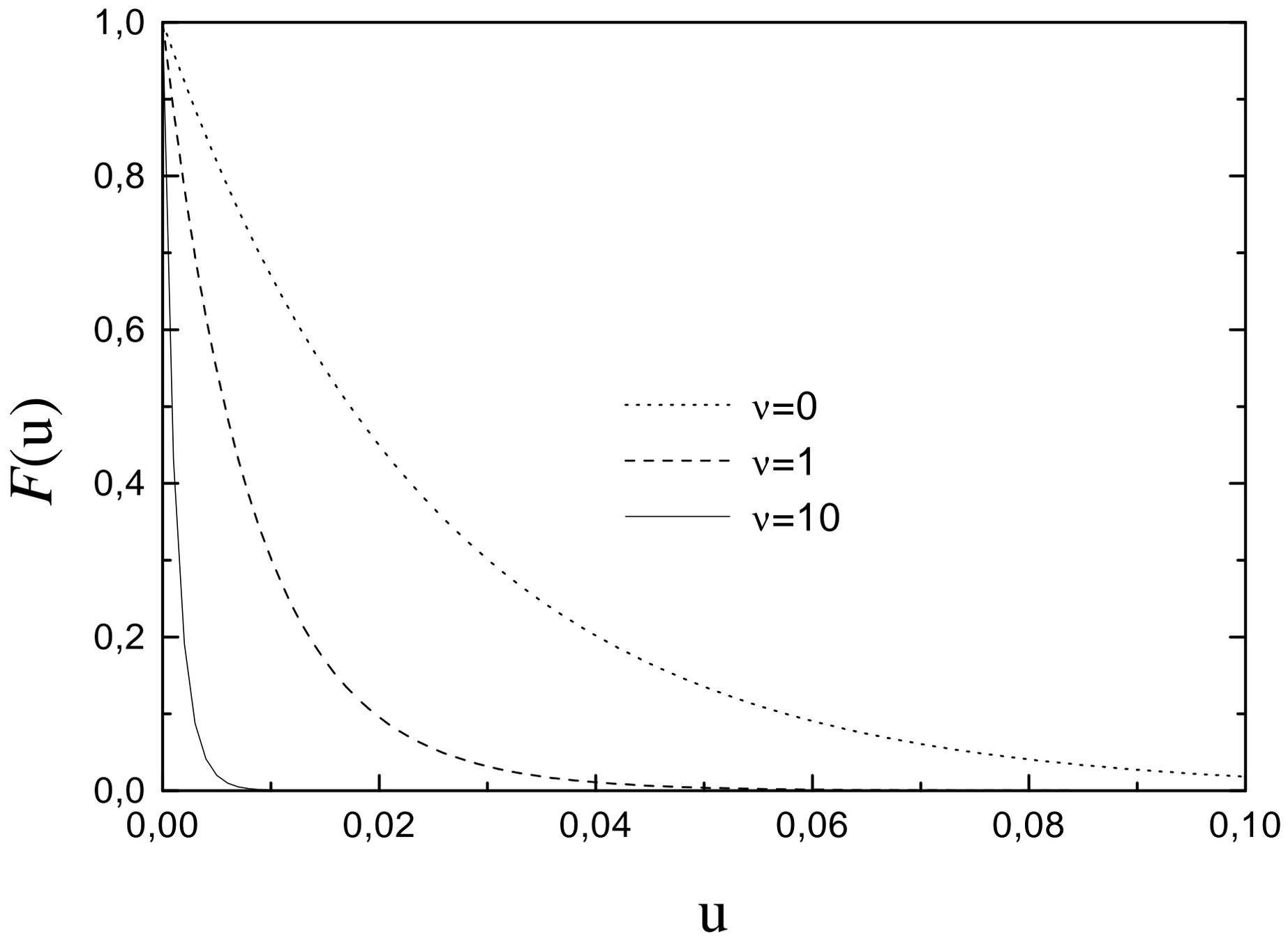
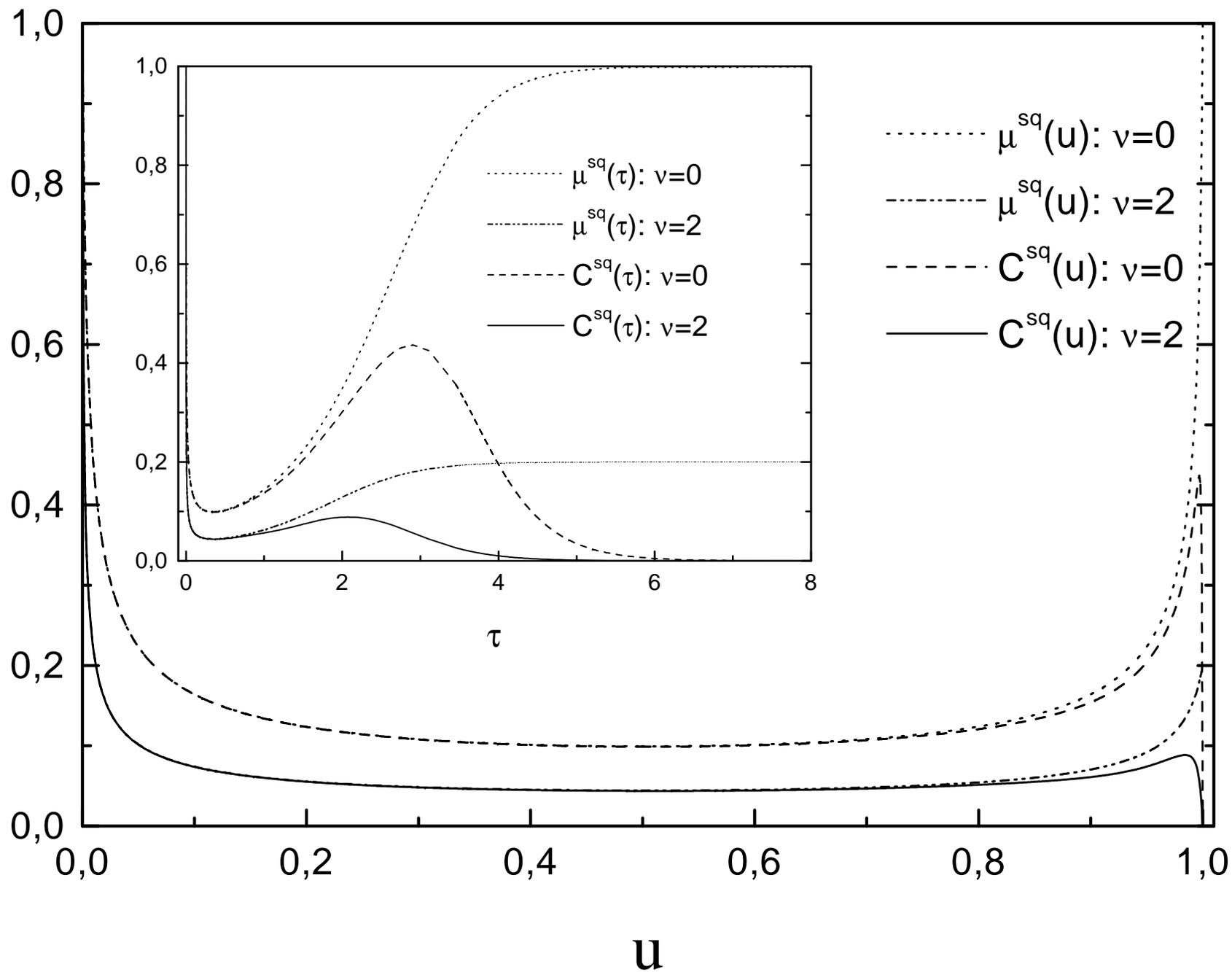
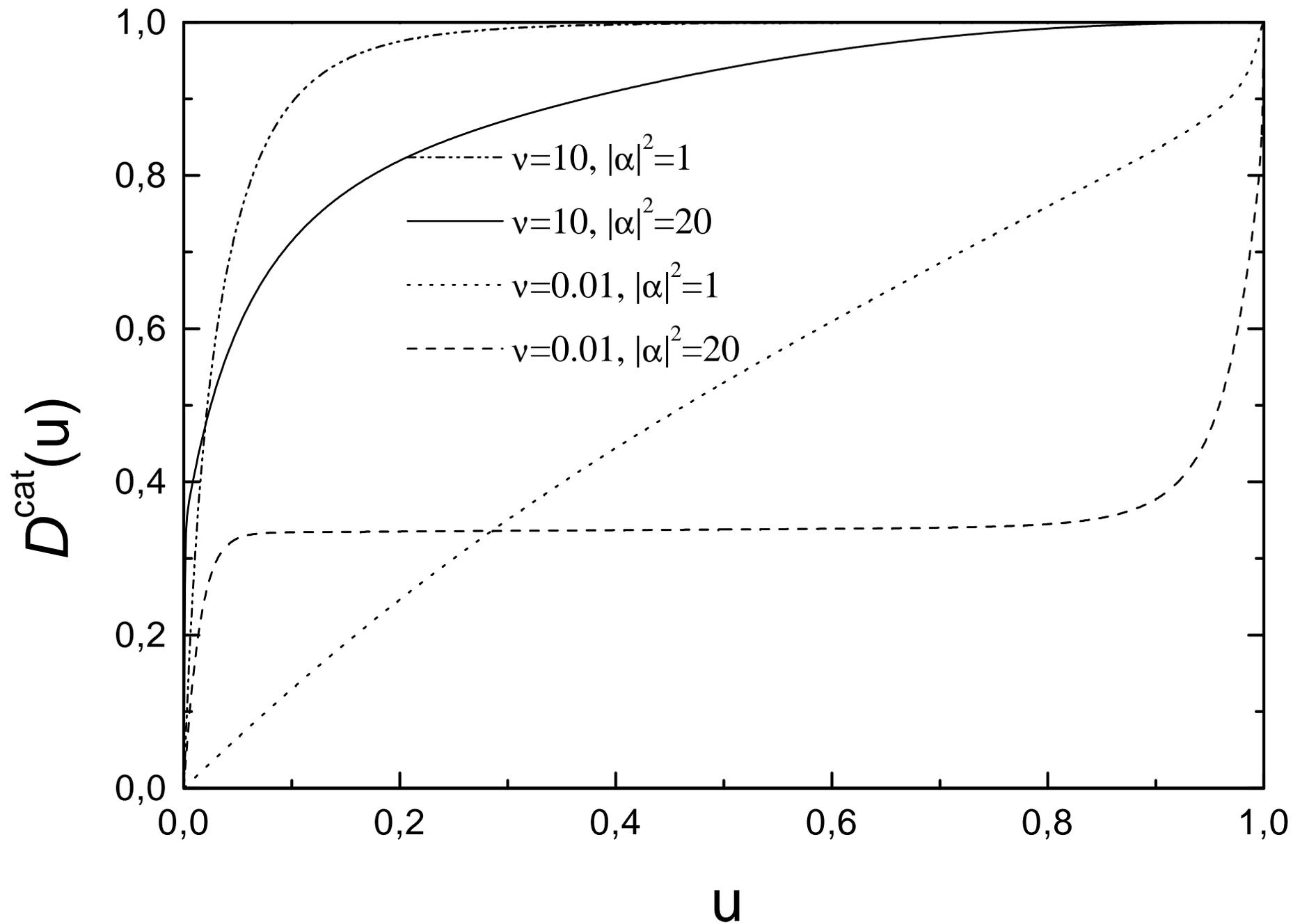
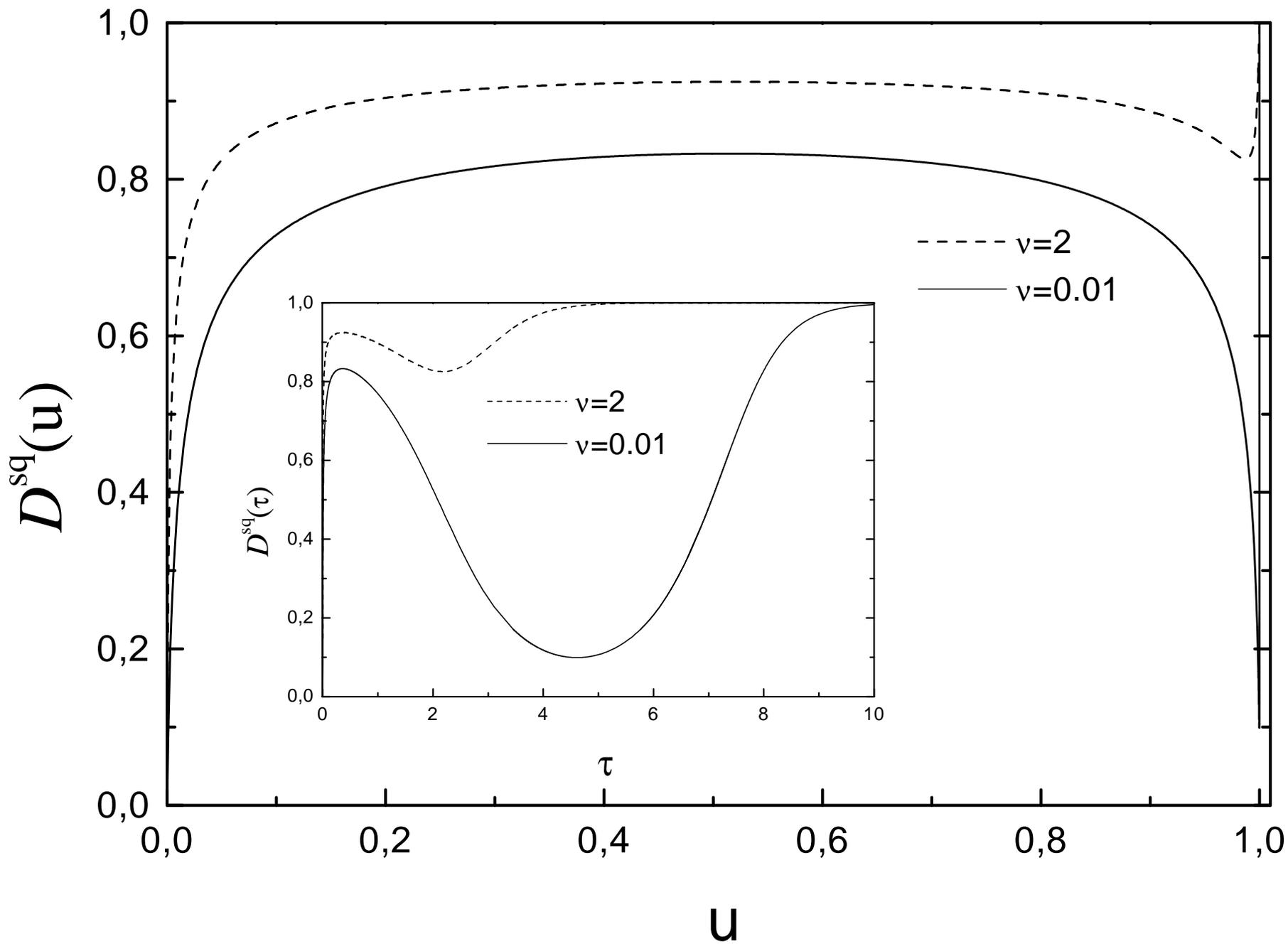
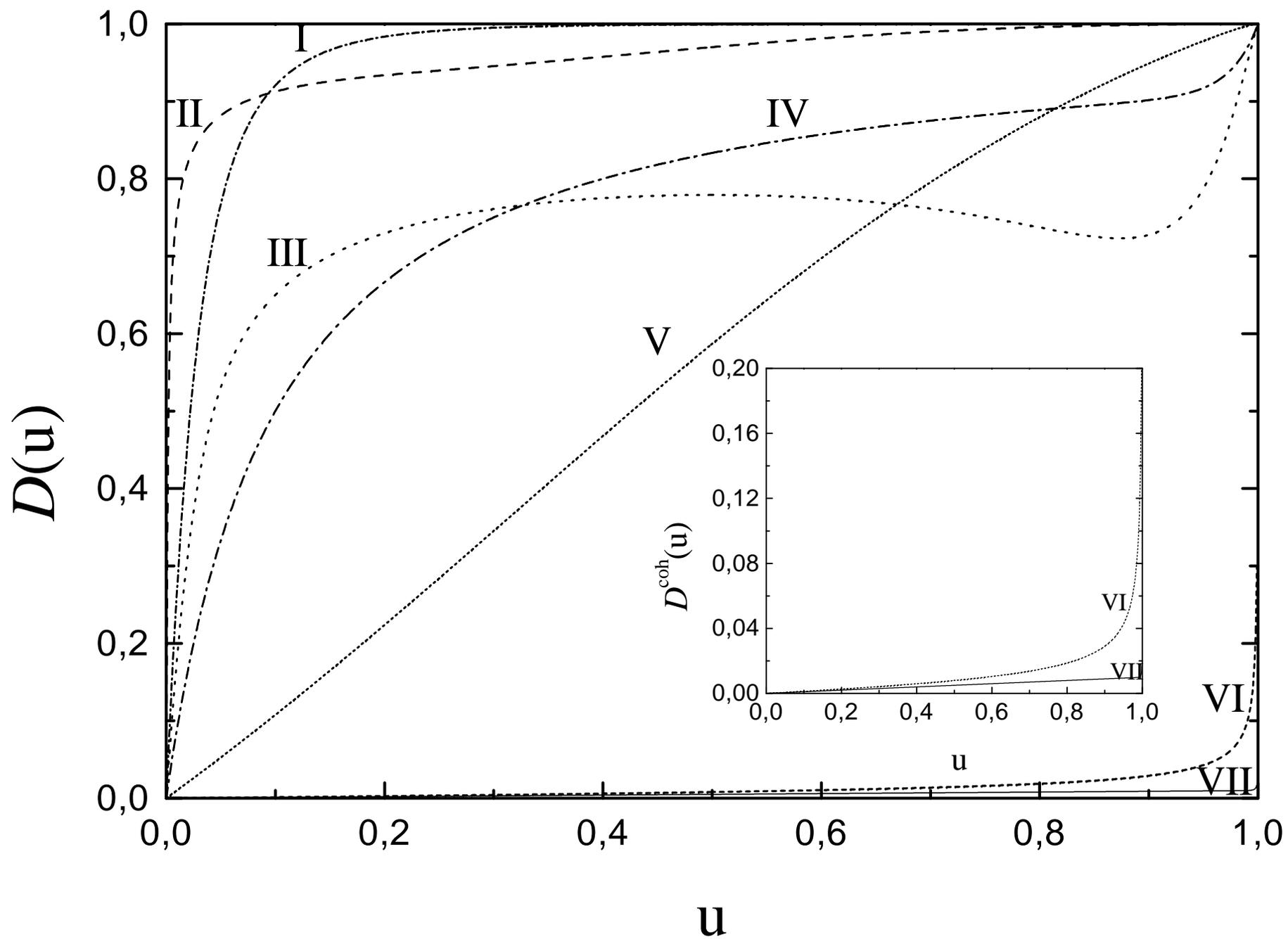

It is interesting to consider also the thermalization of the initial
$M$-photon Fock state $|M\rangle$. In this case
the off-diagonal elements of the statistical operator in the Fock basis are
equal to zero identically for any time $t\ge 0$,
so the `total purity' $\mu$ coincides with the diagonal one $\l$.
The initial diagonal generating function equals $G_0(z)=z^M$.
Consequently,
\[
p_0(t)=\frac{\left[u(t)(1+\nu)\right]^M}
{\left[1+\nu u(t)\right]^{M+1}},
\]
whereas the integral (\ref{calcmud}) can be transformed to the form
\be
\l(t)= \int_0^{2\pi} \frac{d\varphi}{2\pi}
\frac{\left[a + b\cos\varphi\right]^M}
{\left[c -d \cos\varphi\right]^{M+1}}
\label{mu-ru}
\ee
\[
a=u^2(1+\nu)^2+(1-u-u\nu)^2 , \quad  d=2u\nu(1+u\nu)
\]
\[
b=2u(1-u-u\nu) , \quad  c=\nu^2u^2+(1+u\nu)^2
\]
To calculate the integral  (\ref{mu-ru}) we
designate it as $I_M$ and introduce a new generating function
\[
Q(y)=\sum_{n=0}^{\infty}I_n y^n =
\int_0^{2\pi} \frac{d\vf}{2\pi[c-ya -(d+yb)\cos\vf]}
\]
The last integral is given by the expression \cite{Grad}
\[
Q(y)=\left[(c-ya)^2 -(d+yb)^2\right]^{-1/2}
\]
which has the same structure as
the known generating function of the Legendre
polynomials, so after some algebra we obtain
\[
\l=\frac{\left|1-2u(1+\nu)\right|^M}{(1+2u\nu)^{M+1}}
P_M\left(\frac{(1-u)^2 +u^2(1+2\nu)^2}{(1+2u\nu)\left|1-2u(1+\nu)\right|}
\right)
\]
The typical dependences ${\cal D}(u)$ for the coherent and Fock states
with different initial energies are given in figure \ref{fig-8}.

\section{Three stages of decoherence and thermalization}\label{sec-6}

We see that at low temperatures
the decoherence and thermalization of highly excited initial states
go through three distinct stages.
The first one is rather short, its characteristic time being determined
completely by the initial {\it energy of quantum fluctuations\/},
$t_1\sim (\gamma{\cal E})^{-1}$.
However, the coefficients ${\cal C}$ and ${\cal D}$ do not assume their
equilibrium values ($0$ and $1$, respectively) at the end of this stage,
but they remain approximately constant for a rather long period of time.
The total destruction of coherence is observed
only after the time $t_d\sim (2\g)^{-1}\ln\left(E\right)\gg t_1$,
where $E$ is
either the total energy or its `classical' part (depending on the
initial state). This time tends to a finite limit when the temperature $T$
goes to zero.

The disappearance of the off-diagonal matrix elements of the
statistical operator (decoherence) does not mean that the energy level
populations reach their equilibrium values. This happens only after the
`thermalization time' $t_T$, which can be evaluated from the asymptotical
behavior
of the thermalization parameter ${\cal D}$ at $t\to\infty$ in the form of
the Taylor expansion with respect to the small variable 
$\varepsilon =1-u=\exp(-2\g t)$.
For example, for a generic squeezed state with nonzero
mean values of the quadrature components we have (see also (\ref{coh-fast}))
\be
{\cal D}(1-\vep) = 1- \frac{(1+2\nu){\cal E}_{cl}}
{2\nu^2(1+\nu)}\vep  +{\cal O}\left(\vep^2\right)
\label{as-cl}
\ee
where the `classical energy' is given by (\ref{Ecl}).
Assuming (for $\nu\ll 1$)
$\left({\cal E}_{cl}/\nu^2\right)\exp(-2\g t)\sim 1$
we obtain the estimation
$t_T\sim (2\g)^{-1}\ln\left({\cal E}_{cl}/\nu^2\right)$ which shows that
the `thermalization time' may exceed essentially not only the decay
time $\g^{-1}$ but the `ultimate decoherence time' $t_d$, too.
In particular, for the initial coherent state ($\rho=0$) we have
$t_T\sim \g^{-1}\ln|\a/\nu|$.

The situation resembles the classical theory of magnetic relaxation, where
we have also two characteristic times: the time of transverse
relaxation (dephasing)
$T_2$ (analog of $t_d$) and the time of longitudinal relaxation $T_1$
(analog of $t_T$). The difference is that in our case both times
depend not only on the properties of the environment (through the constants
$\g$ and $\nu$), but also on the initial state (through its energy).
Besides, there exists the third time -- the `primary decoherence time'
$t_1$.

For the states with zero mean values of the quadratures, the expansion
of $1-{\cal D}$ begins with the {\it quadratic\/} term
$\vep^2=\exp(-4\gamma t)$.
For the initial {\it vacuum squeezed state\/} ($\a=0$) we have
\be
{\cal D} = 1- \left[\frac{\sinh(2\rho)}{4\nu(1+\nu)}\vep\right]^2
+{\cal O}\left(\vep^3\right).
\label{asDsq}
\ee
If $\rho > 1$ and $\nu\ll 1$, one can rewrite (\ref{asDsq}) as
\be
{\cal D} \approx 1- \left[E\vep/(2\nu)\right]^2
\label{nu2}
\ee
where
$E=\sinh^2\rho +\frac12 \approx \frac14 \exp(2\rho)$
is the total energy of the initial state
(it coincides with the energy of fluctuations in the case involved).
Consequently,
$t_T^{sq-vac}\sim (2\g)^{-1}\ln\left(E/\nu\right)$.

A similar behavior of ${\cal D}(u)$ at $1-u \ll 1$
is observed for the `cat' states. If $a\gg 1$, then the
dependence on the phase $\vf$ becomes unimportant, and we obtain
\[
{\cal D} = 1- \frac{(a\vep)^2 (1+2\nu)}{4\nu^2(1+\nu)^2}
+{\cal O}\left(\vep^3\right).
\]
In this case the total energy $E\approx a \gg 1$, and we arrive again at
the equation (\ref{nu2}) (if $\nu\ll 1$) which yields
$t_T^{cat}\sim (2\g)^{-1}\ln\left(E/\nu\right)$,
 similarly to the case of the vacuum squeezed state.

For the Fock states we obtain
\[
{\cal D}(1-\varepsilon) = 1-\frac{M(M+1)}{4\nu(1+\nu)^2}\varepsilon^2
+\cdots .
\]
If $M\gg 1$, then $E\approx M$, and we have
$t_T^{Fock}\sim \gamma^{-1}\ln (E/\sqrt\nu)$.
We see that the `thermalization time' $t_T$ depends logarithmically on
the initial energy. Besides, it has the strong temperature dependence,
growing as $T^{-1}$ at $T\to 0$ (remember that
$\nu=\left[\exp(\hbar\omega/k_B T)-1\right]^{-1}$, so
$\nu\approx \exp(-\hbar\omega/k_B T)$ at $T\to 0$),
in a complete agreement with the third law of thermodynamics.
The dependence of the `thermalization time' on the mean equilibrium photon
number $\nu$ enables ordering
different families of quantum states with respect to their robustness
against the thermalization (while the `primary time' $t_1$ is the same for
all states with equal values of the energy of quantum fluctuations).
The coherent states are the most robust ones,
then follow squeezed and `cat'
states, whereas the Fock states, being `the most unclassical states',
are thermalized much faster than all the others.

Another interesting feature of the decoherence and thermalization process
is the existence of `plateaus' in the dependences ${\cal C}(u)$ and
${\cal D}(u)$ for several different types of
states (excluding the coherent states) possessing high initial energy
(provided the temperature is low enough).
In the cases of the squeezed and Fock states
the altitudes of `plateaus' tend to $1$ for ${\cal D}(u)$ and to $0$ for
${\cal C}(u)$ when the initial energy increases.
But for the coherent `cat' states,
the metastable values of the coherence and thermalization coefficients
remain finite even for $E\to\infty$: ${\cal C}_{plt}\sim\frac12$ and
${\cal D}_{plt}\sim\frac13$.
Consequently, some degree of coherence (with respect to
the fixed energy basis)
survives in the `cat' states for a long period of time
$t< t_d\sim (2\g)^{-1}\ln a$.
Perhaps, this fact could be important for applications.

\section{Conclusion}\label{sec-7}
We may conclude that the new quantitative measures of decoherence and
thermalization shed new light on the details of the
decoherence process accompanying
the `standard' thermal relaxation of a quantum harmonic oscillator,
showing that this process has three distinct stages in the case of
highly excited initial pure states and low temperatures.
In particular, we have shown that at low
temperatures the `ultimate decoherence' is achieved after rather long
interval of time, which is essentially greater than the relaxation time
and the `primary decoherence time' which was the central subject of
previous studies. Our analysis permits to find the conditions under which
the `purity' or the `linear entropy' can serve as reasonable measures
of (de)coherence: the initial energy of the quantum state must be much
greater then the mean energy of the reservoir oscillators,
$E_0\gg 1+2\nu$. However, even under this condition the `purity' can be
used only to describe the {\em initial\/} stage of the relaxation process,
but it
cannot replace the true measures of `coherence' for the whole time interval.

\section*{Acknowledgements}

ALS thanks CAPES (Brasil) for support.
SSM thanks CNPq (Brasil) for partial financial support.

\appendix
\renewcommand{\theequation}{A.\arabic{equation}}

\section{Propagator of the Fokker-Planck equation}
\setcounter{equation}{0}

Under certain conditions the process of relaxation of {\em linear\/}
$r$-dimensional quantum systems
(such as a system of coupled oscillators or a charged
particle in a homogeneous electromagnetic field and in a confining
parabolic potential) can be described in the framework of
the Fokker-Planck equation for the Wigner function
%\cite{167,DOM85,176,Isar,DOMM95}
\cite{167}, \cite{DOM85}-\cite{DOMM95}
\be
\frac{\partial W}{\partial t}=
 - \frac{\partial }{\partial y_{i}}
\left[ \left({\bf Ay} +{\bf K}\right) _{i }W\right]
+D_{ij}\frac{\partial^2W}{\partial y_i \partial y_j}
\label{FPeq}
\ee
where $ i,j=1,2,\ldots, 2r$;
the $2r$-dimensional vector ${\bf y}$ consists of the linear
combinations of the Cartesian coordinates $q_i$ and canonically conjugated
momenta $p_i$ (in the simplest case ${\bf y}=({\bf q},{\bf p})$).
 The drift matrix ${\bf A}$ and vector ${\bf K}$
 do not depend on the phase space vector variable ${\bf y}$, although they
 may have, in general, arbitrary dependences on time.
However, the diffusion symmetric matrix ${\bf D}\equiv\Vert D_{ij}\Vert$
{\em cannot\/} be arbitrary, since the physically acceptable solutions
to equation (\ref{FPeq}) must satisfy the condition of the positive
semidefiniteness of the corresponding statistical operator. This condition
is fulfilled provided the matrix
${\bf D}_{*}={\bf D}+\frac {i\hbar}4\left({\bf A}\Sigma
+\Sigma\tilde {\bf A}\right)$
is {\em positively semidefinite\/} \cite{167,DOM85,176}.
The elements of the antisymmetric c-number matrix
$\Sigma=\left\Vert\Sigma_{jk}\right\Vert$ are the commutators
$\Sigma_{jk}=\frac i{\hbar}\left[\hat{y}_j,\hat{y}_k\right]$.
In the case of the single space coordinate the matrix condition
${\bf D}_{*}\ge 0$ is equivalent to three scalar conditions
\cite{Zven,Barch,Vals,Sand}
\begin{equation}
D_{pp}D_{qq} -D_{pq}^2\equiv\det
{\bf D}\ge\frac {\hbar^2}{16}(\mbox{Tr}A)^2
\label{176-27}
\end{equation}
\[
D_{pp}\ge 0, \quad D_{qq}\ge 0, \quad
{\bf D}=\left\Vert\begin{array}{cc}
 D_{pp}& D_{pq}\\
 D_{pq}& D_{qq}\end{array}
\right\Vert.
\]
Since (\ref{FPeq}) can be considered as the Schr\"odinger equation with
an effective quadratic (although non-Hermitian) Hamiltonian, the
propagator $G\left( {\bf y},{\bf y}',t\right)$,
\[
W({\bf y},t)=
 \int d{\bf y}^{'}G\left( {\bf y},{\bf y}',t\right)
W\left({\bf y}^{'},0\right),
\]
can be found with the aid of the method of quantum time-dependent
invariants given in \cite{183,75,78}. However, to find its explicit
expression it is sufficient to know that this propagator is a Gaussian,
so, as any Gaussian Wigner function \cite{167,183,Nmod,Teg} it can be
written as
\beqn
&&G({\bf y},{\bf y}',t)=(2\pi )^{-r}\left[\det {\cal M}_{*}
(t)\right]^{-1/2}
\nonumber\\&& \times
\exp\left\{-\frac 12\left[{\bf y}-{\bf y}_{*}(t)\right]
{\cal M}_{*}^{-1}\left[{\bf y}- {\bf y}_{*}(t)\right]\right\}
\label{3.20}
\eeqn
where ${\bf y}_{*}({\bf y}',t)$ is the mean value of the phase space vector
${\bf y}$ and ${\cal M}_{*}(t)$ is the variance matrix.
The explicit form of ${\bf y}_{*}$ and ${\cal M}_{*}$ can be obtained by
solving
the equations (which are immediate consequences of the Fokker-Planck
equation (\ref{FPeq}))
\be
\dot {\cal M}_{*} = {\bf A}{\cal M}_{*} +{\cal M}_{*}\widetilde {\bf A}
+ 2{\bf D}
\label{eqMy}
\ee
\be
\dot{\bf y} = {\bf Ay} +{\bf K}
\label{eqy}
\ee
with the initial conditions
${\cal M}_{*}(0)=0$ and ${\bf y}_{*}({\bf y}',0)={\bf y}'$,
which are equivalent to the property
$G\left( {\bf y},{\bf y}',0\right)=\delta\left({\bf y}-{\bf y}'\right)$
distinguishing the propagator from all other Gaussians.

In the case under study equations (\ref{eqMy}) coincide with the set
(\ref{dqp})-(\ref{equationsqp}). Putting their solutions
(\ref{solq})-(\ref{solsqp}) to the right-hand side of (\ref{3.20})
we obtain the propagator (\ref{prop}).

\newpage
\begin{figure} \caption
{The `purity' $\mu$ and the `coherence' ${\cal C}$
versus the `compact time' $u\equiv 1-e^{-2\gamma t}$ for the initial
coherent state with $|\alpha|^2=1$ and for three diferent temperatures.
The order of the curves from top to bottom (in the left-hand side of the
figure): ${\cal C}$ for $\nu=0$ (zero temperature) (in this case
$\mu\equiv 1$);
$\mu$ for $\nu=1$, ${\cal C}$ for $\nu=1$;
$\mu$ for $\nu=10$, ${\cal C}$ for $\nu=10$.
\label{fig-1}
}
\end{figure}
\begin{figure} \caption
{The `purity' $\mu$ and the `coherence' ${\cal C}$
versus the `compact time' $u\equiv 1-e^{-2\gamma t}$ for the initial
coherent state with $|\alpha|^2=10$ and for three diferent temperatures.
The order of the curves from top to bottom (in the left-hand side of the
figure): ${\cal C}$ for $\nu=0$ (zero temperature) (in this case
$\mu\equiv 1$);
$\mu$ for $\nu=1$, ${\cal C}$ for $\nu=1$;
$\mu$ for $\nu=10$, ${\cal C}$ for $\nu=10$.
\label{fig-2}
}
\end{figure}
\begin{figure} \caption
{The `purity' $\mu$ and the `coherence' ${\cal C}$
versus the `compact time' $u\equiv 1-e^{-2\gamma t}$ for the initial
{\em odd\/} coherent state ($\vf=\pi$) for different values of
the parameters $a\equiv|\alpha|^2$ and $\nu$.
The order of the curves from top to bottom (in the right-hand side of the
figure) is as follows.
I -- $\mu$ for $a=1$ and $\nu=0$;
II -- $\mu$ for $a=10$ and $\nu=0$;
III -- ${\cal C}$ for $a=10$ and $\nu=0$;
IV -- ${\cal C}$ for $a=1$ and $\nu=0$;
V -- $\mu$ for $a=2$ and $\nu=5$;
VI -- ${\cal C}$ for $a=2$ and $\nu=5$.
\label{fig-3}
}
\end{figure}
\begin{figure} \caption
{The `accompanying coherence' ${\cal F}$
versus the `compact time' $u\equiv 1-e^{-2\gamma t}$ for the initial
{\em even\/} coherent state ($\vf=0$) with $|\alpha|^2=10$, for
three different temperatures: $\nu=0,\,1,\,10$.
\label{fig-4}
}
\end{figure}
\begin{figure} \caption
{The `purity' $\mu$ and the `coherence' ${\cal C}$
versus the `compact time' $u\equiv 1-e^{-2\gamma t}$ for the initial
{\em squeezed\/} coherent state with $\rho=3$, $|\alpha|^2=1$ and
$\phi=\pi/2$, for two values of the equilibrium mean photon number in the
reservoir: $\nu=0$ (two close curves relatively far off the bottom)
and $\nu=2$ (two close curves nearby the bottom). In each pair of close
curves the upper one corresponds to $\mu$ while the lower one gives
${\cal C}$. The curves inside the internal box give the same functions
versus the usual (scaled) time $\tau=2\g t$.
\label{fig-5}
}
\end{figure}
\begin{figure} \caption
{The `thermalization coefficient' ${\cal D}$
versus the `compact time' $u\equiv 1-e^{-2\gamma t}$ for the initial
{\em odd\/} coherent state ($\vf=\pi$) with different values of
the parameter $|\alpha|^2=1,\,20$; in the low temperature
($\nu=0.01$) and high temperature ($\nu=10$) cases.
\label{fig-6}
}
\end{figure}
\begin{figure} \caption
{The `thermalization coefficient' ${\cal D}$
versus the `compact time' $u\equiv 1-e^{-2\gamma t}$ for the initial
{\em squeezed\/} coherent state with $\rho=3$, $|\alpha|^2=1$ and
$\phi=\pi/2$, for two values of the equilibrium mean photon number in the
reservoir: $\nu=0.01$ (lower curves) and $\nu=2$ (upper curves).
The curves inside the internal box give the same functions
versus the usual (scaled) time $\tau=2\g t$.
\label{fig-7}
}
\end{figure}
\end{document}